\documentclass[12pt]{article}
 \usepackage{graphicx}
\usepackage{here}
\def\beq{\begin{equation}}
\def\eeq{\end{equation}}
\def\bea{\begin{eqnarray}}
\def\eea{\end{eqnarray}}
\def\bq{\begin{quote}}
\def\eq{\end{quote}}

\def\nin{\noindent}
\def\ba{\begin{array}}
\def\ea{\end{array}}

\begin{document}
\topmargin -1.5cm
\oddsidemargin 0.cm
\evensidemargin -1.0cm
\begin{flushright}
PM/02-61\\
\end{flushright}
\vspace*{1cm}
\begin{center}
\section*{Fundamental Research and Developing Countries \footnote{This an
english version of the communications in french: {\it R\^ole de la Recherche
Fondamentale dans les Pays en voie de D\'eveloppement} (Press: Tribune de
Madagascar n. 4212:  26 November 2002) and {\it Vers une Mondialisation de la Physique
des Hautes Energies} (to appear in Tribune de Madagascar: December
2002).}}
\vspace*{1.0cm}
{\bf Stephan Narison} \\
\vspace{0.3cm}
Laboratoire de Physique Math\'ematique\\
Universit\'e de Montpellier II\\
Place Eug\`ene Bataillon\\
34095 - Montpellier Cedex 05, France\\
\vspace*{1.0cm}
{\bf Abstract} \\
\end{center} 
In the first part of this report, I discuss the sociological r\^ole of fundamental research in
Developing Countries (DC) and how one can eventually
realize this program. In the second part, I give a brief and elementary 
introduction to the field of high-energy physics (HEP), accessible to a large audience not necessary
physicists. The aim of this report is to make politicians and financial backers aware on the long-term
usefulness of fundamental research in DC and on the possible globalisation of HEP and, in general, of science. 
\vspace*{2mm}
\noindent
\section{Introduction}
Fundamental research, like High-Energy Theoretical Physics (HEP), is a missing field in Developing Countries
(DC) like Madagascar, Africa and other thirld world countries. Traditionnally reserved to rich or industrialised countries,
this field of research is now becoming accessible to DC thanks to the progress of communications via Internet, fax 
and the uses of personal computers. However, the most efficient way for realizing this program is a joint collaboration between the DC and
indutrialised countries, where the former can only bring brains and works, while the latter, in addition,  can help for giving funds or/and
fellowships. \\ HEP research consists for studying the constituents of matter and the different laws governing the mechanism of the
universe. In the first part of this report, we shall discuss the sociological r\^ole of the fundamental research in DC. In the second
part, we present in a very simple and pedagogical way (without any formulae!) an introduction to the field of HEP, in the aim that it will
be accessible to a large audience not necessary physicists.
\section{Sociological R\^ole of Fundamental Research in DC}
\subsection{Fundamental research and education}
In most of DC, one has the tendancy to favour the applied and technical formation which is the goal for forming in a
short period the maximum of young peoples for training applied science and technical professions
rapidly exploitable for the country. Though this short term formation is an useful initiative, one should not
also neglect the formation of few young \'elite which will be useful at medium and long-terms in order to
maintain the DC at a high intellectual and technical levels, the most possible way for the DC to reach the level of
industrialised countries. One good possibilty for realizing this high-level formation is the fundamental research. One
can take the example of India and Pakistan which are a rich reserve of researchers despite the
poor economical situation of these countries. Thanks to their traditional philosophy and culture, one can
find their some good scientists known in the world though most of them are expatriate either in Europe or in
the USA. For instance, the physics Nobel price 1979, Pr. Abdus Salam ex-director and founder of the
International Center for Theoretical Physics (ICTP) of Trieste-Italy is a pakistanian. Thanks to the creation
of the ICTP, now named Abdus Salam Center, Pr. Salam has participated in a long term to the development of
science in all DC, because most of thirld world scientists go regularly to this center for recycling and for
getting informed on the last progress of science.\\
This short introduction shows quite well, how scientists can bring a lot to the society. A similar example
should also be applied to the DC, and particularly to Madagascar. Therefore, it is the duty of the political
decision-makers to encourage science in DC. However, this is a medium and long-term objectives, which one can
essentially realize by the improvement of the educational system, which unfortunately has declined during
the last thirty years in Madagascar for various reasons. Indeed, science should belong integrally to our general
culture. Unfortunately, at present, fewer students choose the scientific fields, perhaps, because science needs some
particular effort, or, because our society has not yet understood the r\^ole of science in our culture
and civilisation. Ironically, our society is more and more dependent on the technological developments
issued from the science discovery ( for instance, Internet has been discovered by the
CERN-Geneva physicists (not by Microsoft!) as a tool for communicating experimental data between the different
international teams), but in the same time, the scientific education declines. The russian academician Lev Okun told in
1995 during the HEP international conference in Marseille:\\
{\it More peoples are ignorant, more they hate the spirit of scientific curiosity, and more the processus of
intellectual deterioration is irreversible.}\\
It appears to me that this remark especially applies to the DC, and in particular to Madagascar, when one
observes the majority of young peoples taking the direction towards the study which is useful at short term but often
at a low level. This choice, one should say, is mostly infuenced by the political and
socio-economical environments where these youths are living. In particular, it is the obligation of the
financial backers and of the government, not only to financially support the oriented and applied education
and researches which yield well at short term, but, it is also especially important to encourage the long term
education by providing fellowships and/or training courses. In parallel, they should also take all opportunuities for
popularizing the science and the fundamental research using the modern media (press, radio, television,...)
for communicating to the others the enthousiasm for the science and for explaining its interest. Fundamental
science has a vital r\^ole for this development, because, contrary to applied sciences (ingeneering and technologies),
the researcher is able to form the youths at different scientific and technical levels, from the secondary school to the 
ingeneering studies and PhD scientific diploma. This r\^ole becomes feasible, because physicists have a strong basis
and deep knowledge of sciences (mathematical tools, informatics,...) combined with their talent of researcher or/and
teacher. Indeed, in his profession, a physicist should know how to use these basic knowledges for explaining
observable and measurable natural phenomena. This is an important aspect differentiating a physicist from a
mathematician. For instance, for a mathematician, a sinuso\"\i dal function is $\sin \theta$ or $\cos \theta$, while
for a physicist, this function illustrates different phenomena like the alternate electric current or the wave
propagations...
\subsection{Science and technology: responsability of the scientist}
Most part of our modern society does not even know the traditional goal of science, which is to understand the nature, the
universe and the different laws which govern them. Therefore, it is not surprising that there is a confusion between
science and its utilisation., i.e. between {\it science and technology}, between {\it the knowledge attainments and the
utilisation which the society decides to do with}. When, one talks, for instance, about nuclear physics, most of us thinks
to nuclear power station, nuclear tests or massive destructive bombs, and believe that the science is guilty and socially
irresponsible. When the utilisation of knowledge is an ethical problem, the research of these knowledges should be
absolutely free, and it is advisable not to mix the two things. The preparation of an atomic bomb was due to technology
({\it applied science}), an actual product of ingeneers, though based on the {\it fundamental principles of physics}.
However, the decision to built the hydrogen bomb was {\it political}, but not {\it scientific}. Robert Oppeinhemer, one of
the assumed ``fathers" of the american atomic bomb, explained clearly that:\\
{\it The scientific is not responsible of the natural laws, but his job is to discover how these laws work. However, it is
not the responsability of the scientific to determine if it is suitable or not to use the hydrogen bomb.}\\
The dictates of defense have led the politicians to ask physicists to prepare massive destructive arms. But some other
physicists have also contributed for establishing bridges and links between two adverse countries during the war. The
r\^ole of CERN-Geneva and the ICTP-Trieste on the relations with the former Soviet Union are concrete examples. However,
science should not apologize to the world on what it brings. On the contrary, its contribution represents one of the most
noble and ambitious efforts of the humanity:\\
{\it In trying to discover the nature of the Universe, the scientific attempts to discover himself in this fascinating and
extremely embarrassing world.}\\
{It is the responsability of the scientist to let known to the society the true r\^ole of science,} not only as a
responsability to the taxpayers, but as the basis of the intellectual values. The scientific has the duty to inform the
society, on the existence of tools or methods allowing to analyze complex situations, in order to find an appropriate
solution to the hot problems raising the question of our survival. Therefore, the scientists should inform themseleves on
the nature of important problems to which our society is confronted; in other respects, they should be able to predict the
eventual problems of the future, and inform the society if an appropriate answer already exists. This is indeed their duty
in the crucial domains like the needs of new energy sources, the planet pollution, the decrease of the ozone layer around
the world, the demographic boom, the unproductiveness of the grounds, the decreases of biological species,...But it is
neither the r\^ole nor the responsability of the scientist (this does not mean also that the scientist is not sensitive to
this crucial problem) to resolve the unemployment problem, the famine,...that can only be solved from political decisions.
The {scientist \it r\^ole is at the same time crucial and modest}, in the sense that scientists should not use science for
settling a new type of power, {\it the scientific power}. The scientist should encourage the scientific attitude based on
facts, on curiosity and on audacious search for new fundamental concepts. Michael Faraday has discovered the electricity by
his own intellectual curiosity which guided his fundamental research on the electricity while the society has asked him at
that time to improve the lighting with candles of lighthouses. Reciprocally, the history of sciences has many bitter
examples, when one thinks about the Arab culture of the pre-medieval era, which brought to Europe wide knowledges such as
medicine, mathematics and astronomy. But. for different reasons, this inquistiveness of mind disappeared, letting the
Arab modern society very poor.\\
For this reason, and despite the social context of our country, one should encourage the fundamental and free research
though its technological implications are at a long term, because it is a secure value. It is as expensive as other
cultural values like literature, arts and some other national patrimony, which one should preserved.
\subsection{How to do a fundamental research in DC?}
When I have finished my degree of physics at the University of Antananarivo, I got a fellowship by the European Commussion
to prepare a diploma of ingeneer at the Ecole Centrale or Supelec of Parish. However, when I have hesitated between this
choice and the PhD thesis in Theoretical Physics, my former professor of quantum mechanics Pr. Raoelina Andriambololona has
convinced me by these words:\\
{\it For doing Theoretical Physics, one only needs a pencil and a sheet of paper...}\\
These words perfectly summarize the profession of theoretical physicist in the years of 70 and show the cheap cost of the
formation for this advanced profession, which is then feasible in DC, contrary to that one may a priori think. However, within
the development of the modern society, and for doing fundamental research, one should add to these, the unavoidable
computers as tools for calculations,...and communications (Internet), which hopefully we already have in Madagascar (note
that the insulation was the major handicap for doing a research in our country during the pre-Internet period before 90).
To these tools, one should add the  bi- or multi-lateral {\it scientific cooperations} with advanced
laboratories and institutes, because the exchanges between researchers are necessary at different viewpoint in order to have a
competitive research work at the international level. Though the DC cannot bring funds (which are expected to be obtained from
industrialised countries) for a such cooperation, they can instead  bring brains and works, which are always useful inside a collaboration.
Indeed, these means are relatively cheap and, a priori, feasible for a DC like Madagascar.
\subsection{Can one do HEP in Madagascar?}
Working in advanced domains needs a strong international competitivity. And for being competitive, one needs a good
scientific environment and the possibility to visit regularly big research centers in order to be informed on new
experimental data and new theoretical progresses. Unfortunately our country, was an inauspucious place (geographical and
economical situations), and then, you can understand immediately, why I did not come back to Madagascar, despite my deep
wishes to serve my nation. However, this choice of foreign residence should not lead to a total desinterest to the country
but on the contrary should stimulate finding a way to contribute to its recovery. This is the reason why I have decided to
create the HEP-MAD institute, project one which we have started to work in collaboration with the national teachers and
researchers since last year. This institute will be useful for forming young students at high level and will serve as a
platform to different international cooperations and HEP-MAD conferences. We have initiated the latter in september 2001 (HEP-MAD 01
conference).\\ We wish sharply that the new governement and the financial bakers give all means for concretising rapidly this project
because it goes along the lines of their program for render Madagascar a competitive country at the international level.
\section{Towards a globalisation of HEP}
In this second part, we try to vulgarize this advanced research of HEP by limiting ourselves to the basic concepts and
illustrative explanations in the aim to present a very elementary, pedagogical and qualitative report accesible to the majority of
the population who are not necessarily physicists.
\subsection{The structure of matter}
\begin{figure}[hbt]
\begin{center}
\includegraphics[width=17cm]{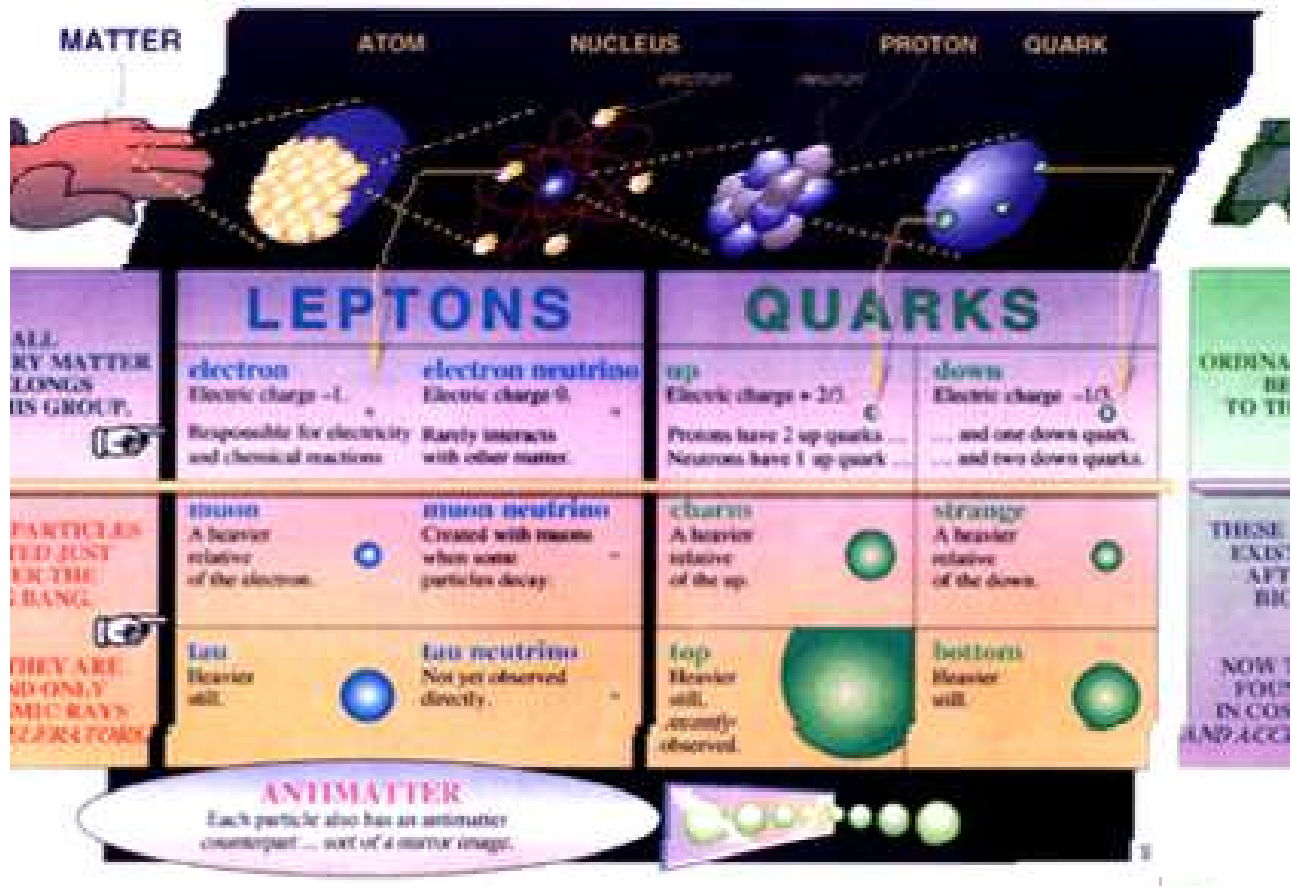}
\caption{Exploration of matter: from macroscopic to elementary particles.}
\label{fig:main}
\end{center}
\end{figure}

In Fig. \ref{fig:main}, one explains, from the
example of hand (matter), the exploration of the underlying structure of matter before reaching the fundamental constituents (elementary
particles) named leptons and quarks.

\nin
Imagine that you have started at the university to study the macromolecules of ADN (Acid
DesoxyriboNucleic) in {\it biochemistry}. After, you go to the degree in chemistry ({\it organic chemistry})
where you study the properties of desoxyribose and of benzen molecules and their by-products. You
continue your way and, with more energy, you succeed to bring the molecule into an atom of carbon (
Avogadro and Gay-Lussac laws). With a more energy, you can break the atom for studying its nucleus and the
electrons which orbit around it ({\it atomic physics}). You continue your breakage and you will
discover that the nucleus is composed by {\it protons and neutrons} linked together by the nuclear
force ({\it nuclear physics}). Finally, bombing the proton, you will find that it is made by elementary
particles which are quarks glued together by gluons ({\it HEP area}). The particle accelerators are
conceived that you can do in practice the exploration of matter thanks to the energy which these accelerators
communicate to particles. As you can see in Fig. \ref{fig:main}:
\begin{itemize}
\item {\bf The electron} is the 1st elementary particle known since a century. It is the 1st member of
the particle family called {\it leptons}, where the other members differ from the electron by their
masses (the {\it muon} and the ${tau}$ are respectively 200 and 3600 times heavier than the electron). To these three
particles of electric charge -1, are associated electric neutral particles ({\it neutrinos}) lighter
and which do not interact with matter. These particles existed after the Big Bang and are produced by
particle accelerators or cosmic rays.
\item {\bf The proton and neutron} are composed by elementary particles called quarks which have
fractional electric charge and possess three colours. The quarks are also consituted of three couples
($u$p, $d$own), ($c$harm, $s$trange) and ($b$eauty, $t$op). Each quark differs by their masses while in each couple
(...,...), quarks have respectively electric charges 2/3 ({\it u, c, t}) and -1/3 ({\it d, s, b}).
Then, a proton is constituted by two quarks $u$ and one quark $d$ ($uud$) which the sum of charge is +1. The
neutron is neutral and composed by $ddu$...
\item To these particles are associated {\it anti-particles} which form the {\it anti-matter}.
\item The repartition of charges and the correspondence between lepton and quark families are essential
for having mathematical consistent theories (renormalizability,...)
\end{itemize}
\subsection{The fundamental forces of nature and the HEP theories}
Nature is gouverned by four fundamental forces, which by order of decreasing range are classified as:
\begin{itemize}
\item {\bf The strong interaction nuclear force} mediated by eight coloured massless gluons responsible for the
cohesion of quarks inside the nucleus has a range of $10^{-12}$ cm. To this force and to the properties of quarks
is associated the {\it theory of Quantum ChromoDynamics (QCD)}.
\item {\bf The electromagnetic force} mediated by the photon is responsible of the light and has a range $10^-3$
weaker than the previous one. To this force and to the properties of leptons is associated the {\it theory of Quantum
ElectroDynamics (QED)} tested, at present, with very high-precision.
\item {\bf The weak interaction force} mediated by the gauge bosons having masses 160 000 times the one of the
elctron is responsible for the $\beta$ radiactivity and has a short range of about $10^{-17}$ cm. To this force is
associated the {\it theory of Weak Interactions}.
\item {\bf The gravitational force} eventually mediated by the gravitons has a macroscopic range of about
$10^{-50}$ cm and is responsible of our weight (recall the story of the famous Newton's apple). Its corresponds to
the {\it theory of Gravitation}.
\end{itemize}
QED and the theory of weak interactions, which, a priori, correspond to completely different forces are put
together in a unique theory called commonly the {\it Standard Model of Electroweak Interactions (SM)}. The discovery of
the SM has led to the attribution of a Nobel price to their discoverers: Prs. S. Glashow (Harvard Univ.-USA),
S. Weinberg (Austin Univ.-USA) and A. Salam (ICTP-Trieste-Italie), while more, recently, the physics Nobel
price 1998 has been attributed to Prs. G. 't Hooft and M. Veltman of the Utrecht Univ-Holland, who have
proved the mathematical consistencies of the SM, rendering it to be a theory not only a phenomenological
model. QCD belongs to the strong link of the the SM, where the theory possesses the asymptotic freedom
property at high-energies rendering easier different approximate perturbative calculations within QCD. There
also exist some {\it Grand Unified Theories (GUTS)} like for instance {\it supersymmetry} which can unify these three
forces to one force at higher energies. In fact, the goal of HE Physicists is to understand the laws governing the
universe within a simple and elegant theory. The search for an unified and mathematically consistent theory,
including the gravitational force is the present challenge of HE Physicists.
\subsection{The particle accelerators}
\begin{figure}[hbt]
\begin{center}
\includegraphics[width=6cm]{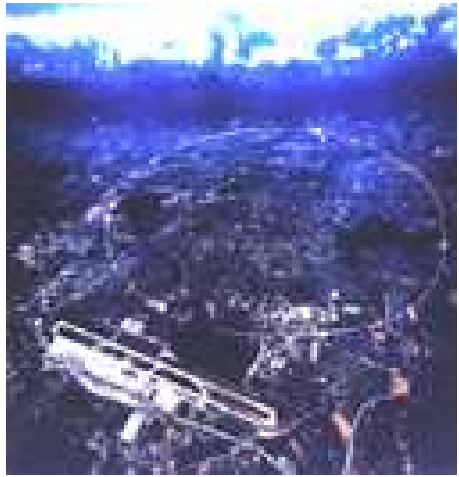}
\caption{Aerial view of the LEP underground tunnel of CERN-Geneva.}
\label{fig:lep}
\end{center}
\end{figure}
\begin{figure}[hbt]
\begin{center}
\includegraphics[width=6cm]{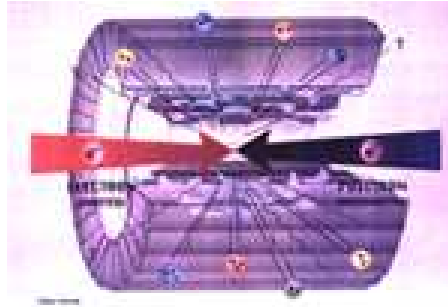}
\caption{Schematic view of the LEP-detector after the electron-positron collision.}
\label{fig:detecteur}
\end{center}
\end{figure}

\nin
Today, science has important means for supporting advanced fundamental research, in particular, in the physics of
space (Voyager, Hubble telescop,...) and in the field of HEP (which I know better) for studying the fundamental
constituents of matter (quarks, leptons, gauge bosons,...) and their r\^ole in the birth of the universe. These
domains need gigantic instruments, that a unique country cannot finance, like the large electron-positron collisionner
LEP (and in the near future the Large hadron Collider (LHC)) at CERN-Geneva, in an underground tunnel of 9km diameter,
straddling the France-Switzerland border. One can see in Fig. \ref{fig:lep}, an aerial view of CERN and the LEP tunnel
where the airport of Geneva is visible in front. \\
In Fig. \ref{fig:detecteur}, one shows the reaction inside the LEP detector after the collision of the electron with its
anti-particle, the positron, where new heavier particles are produced.

\nin
Though apparently expensive, these experiments are far from being useless, as these are material and intellectual
tools which allow the humanity to make up all challenges to which it is, at present, confronted.
\subsection{What are the socio-economical implications of HEP?}
Apparently,       this fundamental
research is       far from the every days life applications.       However,
this is not really true as       any applied science and technologies
      need results from       fundamental research.       There are many
indirect consequences       of this research:
\begin{itemize}
      \item Electricity       which we use everyday comes from the discovery of the      
electron  (1st elementary particle)       and of
its properties.
      \item TV screen       is the smallest accelerator of particle (electron). It is a mini-LEP where LEP energy is about 5$\times 10^6$
the one of the TV. The similarity of the technologies behind the TV and LEP is commented in Fig. \ref{fig:tv2}.
\begin{figure}[hbt]
\begin{center}
\includegraphics[width=10cm]{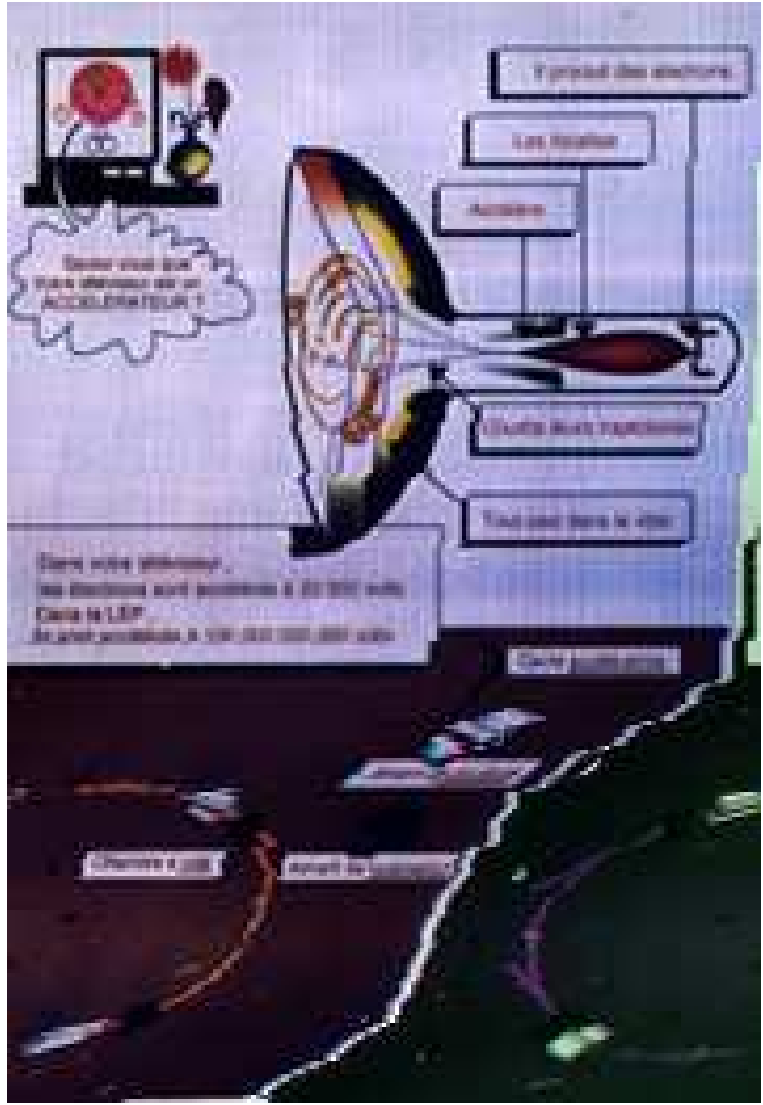}
\caption{A schematic comparison of the technologies behind the TV and LEP.}
\label{fig:tv2}
\end{center}
\end{figure}
\nin     
 \item Detector buildings  stimulate high-technology in industry and in other
branches of physics (semi-conductors,...) ;       Charpak       type-detectors are used in the
container-custom controls (le Havre),... 
     \item Accelerator technology is used in       medicine       for e.g. the cancer treatment
      (Orsay,...)...
     \item Internet Web pages were  discovered at CERN (Geneva) being originally used for
communicating data and informations between members of experimental groups.
      \item The treatment of HEP data needs powerful computers.
      \item       {\bf ...}
\end{itemize}
\section{Summary and conclusions}
In this short report, I have presented the social r\^ole of fundamental research in DC by emphasizing on its long-term usefulness
and on the possible realization of the project in DC. I have also given a very elementary introduction to high-energy physics at
the level accessible to a majority of readers not necessarily physicists. The aim of this report was to make politicians and financial
backers aware on the long-term usefulness of the fundamental research in DC. I wish that after reading it, they have been convinced on the
feasability of the project in DC.

\end{document}